\documentclass[conference]{IEEEtran}
\IEEEoverridecommandlockouts
\usepackage{cite}
\usepackage{amssymb}
\usepackage{amsmath,amsfonts,amsthm,bm}
\usepackage{mathtools}
\usepackage[linesnumbered,ruled, vlined]{algorithm2e}
\usepackage{graphicx}
\usepackage{textcomp}
\usepackage{xcolor}
\usepackage{comment}
\usepackage{dsfont}
\graphicspath{{images/}}

\newcommand{\RNum}[1]{\uppercase\expandafter{\romannumeral #1\relax}}
\def\BibTeX{{\rm B\kern-.05em{\sc i\kern-.025em b}\kern-.08em
    T\kern-.1667em\lower.7ex\hbox{E}\kern-.125emX}}

\usepackage[nowarn,acronyms,nonumberlist,nopostdot,nomain,nogroupskip]{glossaries}
\newacronym{ra}{RA}{random access}
\newacronym{rl}{RL}{reinforcement learning}
\newacronym{aop}{AoP}{age of packet}
\newacronym{aoi}{AoI}{age of information}
\newacronym{eb}{EB}{exponential backoff}
\newacronym{beb}{BEB}{binary exponential backoff}
\newacronym{dqn}{DQN}{deep Q-network}
\newacronym{fs}{FS}{fixed scheduling}
\newacronym{ed}{ED}{event drive}
\newacronym{mtc}{MTC}{machine-type communication}
\newacronym{mtd}{MTD}{machine-type device}
\newacronym{bs}{BS}{base station}
\newacronym{htc}{HTC}{human-type communication}
\newacronym{drl}{DRL}{deep reinforcement learning}

\begin{document}

\title{Collision Resolution with Deep Reinforcement Learning for Random Access in Machine-Type Communication\\

}

\author{\IEEEauthorblockN{Muhammad Awais Jadoon,
Adriano Pastore, and
Monica Navarro,}
\IEEEauthorblockA{Centre Tecnològic Telecomunicacions Catalunya (CTTC)/CERCA, Castelldefels, Spain}

Email: \{mjadoon, apastore, mnavarro\}@cttc.es
}


\maketitle

\begin{abstract}
Grant-free \gls{ra} techniques are suitable for \gls{mtc} networks but they need to be adaptive to the \gls{mtc} traffic, which is different from the human-type communication. Conventional \gls{ra} protocols such as \gls{eb} schemes for slotted-ALOHA suffer from a high number of collisions and they are not directly applicable to the \gls{mtc} traffic models. In this work, we propose to use multi-agent \gls{dqn} with parameter sharing to find a single policy applied to all \glspl{mtd} in the network to resolve collisions. Moreover, we consider binary broadcast feedback common to all devices to reduce signalling overhead. We compare the performance of our proposed DQN-RA scheme with \gls{eb} schemes for up to 500 \glspl{mtd} and show that the proposed scheme outperforms \gls{eb} policies and provides a better balance between throughput, delay and collision rate.
\end{abstract}

\begin{IEEEkeywords}
Random access, multi-agent DRL, MTC, packet delay, collision resolution.
\end{IEEEkeywords}

\section{Introduction}
The \Gls{mtc} paradigm poses multiple challenges in terms of multiple access due to different its traffic characteristics as compared to the conventional \gls{htc}. For \gls{mtc}, a small area is expected to have a thousand number of low-power low-complexity \glspl{mtd} with different sleep cycles and having short packet length \cite{bockelmann16}. To manage massive access in such a scenario, if on the one hand grant-based scheduling techniques incur a huge signaling overhead, on the other hand, uncoordinated grant-free \gls{ra} schemes that are more favorable for \gls{mtc} traffic such as \gls{eb} schemes for slotted ALOHA, suffer from high number of collisions. In uncoordinated \gls{ra}, each device selects a random physical resource and transmit its data to the receiver. A huge amount of work on \gls{eb} schemes in the literature exists but their performance is highly dependent on traffic arrival models and due to the complexity of the process involved, the analytical approaches provide different solutions for varying underlying assumptions \cite{Barletta18}. 

In the recent state-of-the-art works, \gls{rl}-based schemes for multiple access are popular due to their ability to adapt to different traffic models. Several works including \cite{CHU201523, Naparstek, zhong2019deep, wang2018, Tomovic20, alfaro_alohaq} but not limited to, have proposed \gls{drl}-based \gls{ra} solutions in wireless networks. However, their solutions are not tailored for \gls{mtc} networks and most of these works consider all devices to be active and always having a packet in their buffer. This is not the case for \gls{mtc} networks where due to the varying sleep cycles of battery-constrained \glspl{mtd}, they can become active/inactive in the network randomly. Moreover, for multi-agent \gls{drl}, these schemes do not provide insights into whether their proposed schemes are scalable to a higher number of users in the network or not, when a single resource (channel) is shared among users. The methods proposed in \cite{Yang21} still incur a high signaling overhead for scheduling and also for cooperation among devices.

In our previous work \cite{jadoon2022deep}, we proposed \gls{dqn}-based \gls{ra} algorithm a (DQN-RA) that adapts to different packet traffic arrival rates that follow independent Poisson processes and it provides better performance terms of throughput and fairness as compared to the \gls{eb} schemes. Our proposed DQN-RA scheme is not dependent on any specific arrival process and therefore, any random arrival process can be employed. In this work we extend our model to a higher number of devices and show how \glspl{mtd} cooperatively resolve collisions and empty their packet buffers within $K$ time slots, where $K$ is variable and it is dependent on the total number of \glspl{mtd} in the network and the traffic arrival rate. Moreover, we consider that \glspl{mtd} can become active/inactive in the network and even new devices can join the network. To reduce the signaling, we consider binary broadcast feedback that informs all the active devices whether or not a collision has occurred at each time slot instead of the feedback sent to each device individually.
\begin{figure}
    \centering
    \includegraphics[width=230px, height=90px]{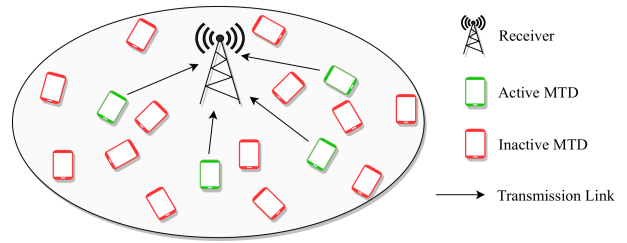}
    \caption{An \gls{mtc} network with active and inactive MTDs and a common receiver} \label{fig:mtc_network}
\end{figure}
\section{System Model and Problem Formulation}\label{sec:system_model}
We consider a synchronous time-slotted \gls{mtc} network as shown in Fig. \ref{fig:mtc_network} with a set $\mathcal{N} = \{1, \dotso, N \}$ of \glspl{mtd} and an error-free broadcast channel. The physical time is divided into discrete slots, each of duration normalized to $1$ and the slot index is $k \in \mathbb{N}$. We assume $\mathcal{N}_a(k) \subseteq \mathcal{N}$ as the set of active \glspl{mtd} at time slot $k$. The packet arrival follows a Poisson process for each device $n$ with average arrival rate $\lambda_n$. We assume the device $n$ to be active when it has a packet in its buffer $B_n(k)$ at time slot $k$, otherwise it is assumed to be inactive, i.e., $B_n(k) \in \{0,1\}$. We assume that each device $n$ can only store a maximum of one packet in its buffer. Furthermore, we assume that at each time slot $k$, an \gls{mtd} can transmit only one packet. \glspl{mtd} are assumed to be slot-synchronized. After transmitting the packet, the \gls{mtd} goes back to inactive/sleep mode. At each time slot $k$, if  $B_n(k)=1$, the device $n$ takes an action $A_n(k) \in \mathcal{A} = \{0,1\}$, where $A_n(k)=0$ corresponds to the event when the device $n$ chooses to not transmit and $A_n(k)=1$ corresponds to the event when device $n$ transmits a packet on the channel. Furthermore, for the inactive devices, we let $A_n(k)=0$. After taking an action, we assume that for each time slot $k$, the receiver sends a broadcast feedback signal $F(k)$ to all the active devices. We define 
\begin{align}\label{eq:broadcast_feedback}
    F(k) = 
    \begin{cases}
           0 & \text{if there is a collision} \\
           1 & \text{otherwise.}
    \end{cases}
\end{align}
The feedback signal $F(k)$ and the action $A_n(k)$ of each \gls{mtd} can be used to calculate the \emph{collision} event $C_n(k) \in \{0,1\}$ and the \emph{success} event $G_n(k) \in \{0,1\}$ for each device $n$, i.e., $g:(F(k), A_n(k)) \mapsto \big(C_n(k), G_n(k)\big)$. These events are locally computed by each  \gls{mtd} and define the success event $G_n(k)$ when a packet has been successfully transmitted by the device $n$ as 
\begin{align} \label{eq:success_event}
        G_n(k) &= 
        \begin{cases}
    		1 & \text{if $A_n(k) = 1$ and $F(k)=1$}\\
    		0 & \text{otherwise}.
    	\end{cases} 
    \end{align}
Similarly, the collision event $C_n(k)$ happens when two or more packets collide with each other and we define $C_n(k)$ as,
\begin{align} \label{eq:collision_event}
        C_n(k) &= 
        \begin{cases}
    		1 & \text{if $A_n(k) = 1$ and $F(k)=0$}\\
    		0 & \text{otherwise.}
    	\end{cases} 
\end{align}
The collided packets need to be retransmitted until they are successfully received at the receiver using the proposed collision resolution scheme calculated as in Section \ref{sec:rl_env}. 

Moreover, we assume that each \gls{mtd} keeps a record of its previous actions, feedback and its current buffer state $B_n(k)$ up to $h$ past instants, $h$ being the \emph{history size}. Hence, at each time slot $k$, the tuple
\begin{align}\label{eq:history_user_n}
    S_n(k) & =  \big( A_n(k-h), F_n(k-h), A_n(k-h-1), \\
    & F_n(k-h-1), \dotso, A_n(k-1), F_n(k-1), B_n(k)\big) \nonumber 
\end{align}
is referred to as the \emph{local history} or the \emph{state} of device $n$, and  $S_{\mathrm{H}}(k) = \big(S_1(k),\dotsc,S_N(k)\big) $
is the \emph{global history} of the system.

In this work, we  are interested in developing a distributed transmission policy $\pi(\cdot)$ for slotted RA that can effectively resolve collisions without excess packet delay and also provide better throughput.
We can mathematically define our objective function as
\begin{align}
    \max_{A_n(k)} \quad & \sum_{k=1}^K G_n(k) - \rho C_n(k),  \forall n \in \mathcal{N},
\end{align}
where $\rho$ is the weightage given to the collision by device $n$. To achieve this objective, we use \gls{dqn} algorithm with parameter sharing for our multiagent/multiuser environment presented in Section. \ref{sec:rl_env}.

\subsection{Performance Metrics}
\subsubsection{Throughput}
The average packet success rate or throughput is defined as the number of successfully delivered packets till the total time $K$. We define the average throughput as,
\begin{align}
    T = \frac{1}{K}\sum_{k=1}^K \sum_{n} G_n(k).
\end{align}
\subsubsection{Packet Collision Rate}
We define the packet collision rate as the number of times collision events happened over time $K$. The average collision rate is therefore defined as,
\begin{align}
    Z = \frac{1}{K}\sum_{k=1}^K F(k).
\end{align}

\subsubsection{Packet Delay}
The delay $D_n(i)$ of the $i$-th packet that has entered the buffer of  \gls{mtd} $n$, is defined as the number of time slots between its entrance into the buffer, and its successful transmission. The total number of packets that have been successful transmitted by device $n$ within $K$ time slots, is $\sum_{k=1}^K G_n(k)$. The sum of delays is equal to
\begin{equation}
    \sum_i D_n(i)
    = \sum_{k} \mathds{1}\{B_n(k)>0\},
\end{equation}
where $\mathds{1}\{\cdot\}$ is the indicator function. We calculate the average packet delay for device $n$ after $K$ time slots as
\begin{align}\label{eq:delay_n}
    d_n = \frac{\sum_i D_n(i)}{\sum_{k=1}^K G_n(k)},
\end{align}
and the average delay for the whole system is $\mathcal{D}= \frac{1}{N}\sum_n d_n$

\subsection{Baseline Exponential Backoff Policies}
We consider \gls{eb} policies as our baseline schemes with backoff factor of $\sigma$. In this paper, we divide \gls{eb} schemes into non-symmetric \gls{eb} (nSEB) and symmetric \gls{eb} (SEB). 

\subsubsection{Non Symmetric Exponential Backoff Policy}
Let us denote the transmit probability of  \gls{mtd} $n$ at time slot $k$ as $p_n(k)$. For \gls{eb} schemes, the transmit probability of the device $n$ after $j$ consecutive collisions becomes $p_n(k) = \sigma^{-j}$. For non-symmetric \gls{eb}, if $i \in \mathcal{N}_c(k)$ where $\mathcal{N}_c(k) \subseteq \mathcal{N}_a(k)$ is the set of colliding  \glspl{mtd}, then the transmit probability of colliding  \glspl{mtd} $i \in \mathcal{N}_c(k)$ can be written as,
\begin{align}
    p_i(k) &=
    \begin{cases}\label{eq:nseb}
        \max\big(\frac{p_i(k-1)}{\sigma}, p_{\mathrm{min}} \big) & \text{if collision}\\
        p_{\mathrm{max}} & \text{otherwise,}
    \end{cases}
\end{align}
where $p_{\mathrm{max}}$ and $p_{\mathrm{min}}$ denote the maximum and minimum transmit probabilities respectively. For $\sigma = 2$, the scheme is referred to as binary nSEB (BnSEB). The Equation \eqref{eq:nseb} shows that the transmit probability is reduced by colliding  \glspl{mtd} only when a collision event occurs. BnSEB is a standard \gls{eb} scheme that has been used in IEEE 802.11 and IEEE 802.3 standards.

\subsubsection{Symmetric Exponential Backoff Policy}
In symmetric \gls{eb} (SEB), each active  \gls{mtd} increases or decreases its transmission probability $p_n(k)$ whenever a collision or a no-collision event happens, respectively. Since all devices have the same transmit probability we drop the subscript $n$ and denote it as $p(k)$, which is calculated as
\begin{align}
    p(k) &=
    \begin{cases}\label{eq:seb}
        \max\big(\frac{p(k-1)}{\sigma}, p_{\mathrm{min}} \big) & \text{if collision}\\
        \min\big(\sigma p(k-1), p_{\mathrm{max}}\big) & \text{otherwise}
    \end{cases}
\end{align}
For $\sigma=2$, the scheme is referred to as binary SEB (BSEB).

\section{\gls{rl} Environment and Multiagent DQN}\label{sec:rl_env}
\subsection{RL Environment Formulation}
\begin{figure}
    \centering
    \includegraphics[width=250px, height=110px]{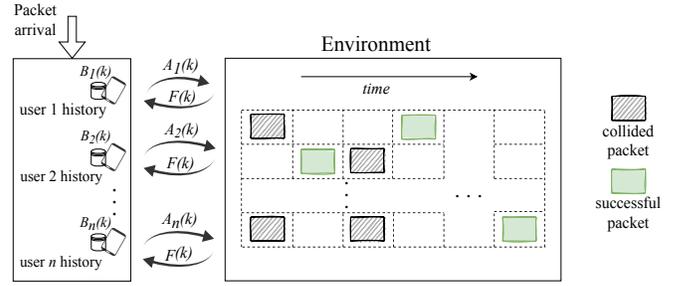}
    \caption{Interaction of agents/devices with the proposed environment. For each time slot $k$, each active device $n$ takes an action $A_n(k) \in \mathcal{A}$, and receives the feedback signal $F(k)$. The devices update their buffer state $B_n(k)$ depending on the feedback signal and the action taken} \label{fig:MURL_env2}
\end{figure}

The define our environment as the multi-agent environment with each \gls{mtd} as an agent\footnote{We use the terms \gls{mtd}, device and agent interchangeably in the rest of the paper.}, and the physical resource (channel) is shared by all the agents as shown in Fig. \ref{fig:MURL_env2}. The environment is partially observable because each agent $n$ is unaware of the actions of other agents and it takes it own action $A_n$ independently based on the observed state $S_n$.
\subsubsection{State} 
We define the \emph{state} of each agent as the local history $S_n(k)$ observed by the agent at each time slot $k$.\footnote{The terms \emph{history} and \textit{state} are used interchangeably in the rest of the paper.} \subsubsection{Actions} 
Similarly, the action of each agent as mentioned above is to transmit $A_n(k)=1$, or to wait $A_n(k)=0$.

The state $S_n$ of inactive \glspl{mtd} is masked with zeros and the corresponding action $A_n$ value is also set to zero.

\subsubsection{Reward}
In \gls{rl}, the goal of an agent in \gls{rl} is to maximize the long-term expected reward and therefore, the reward function reflects the optimization goal for the environment.
Let $R_n(k) \in \mathbb{R}$ be the \emph{immediate} reward that agent $n$ obtains at the end of time slot $k$ after taking the action $A_n(k)$ and receiving the observation $F(k)$ from the environment. The accumulated discounted reward for agent $n$ is defined as
\begin{align}
    \mathcal{R}_n(k) = \sum\limits_{k'=0}^{\infty}\gamma^{k'} R_n(k+k'+1),
\end{align}
where $\gamma \in (0,1]$ is a discount factor.
For our system model, we define the reward $R_n(k)$ as,
\begin{align}\label{eq:reward_c}
    R_n(k) = \sum_n G_n(k) - \rho C_n(k),
\end{align}
The summation sign in \eqref{eq:reward_c} shows that the reward  is global, i.e., all  \glspl{mtd} share the same reward, which indicates that the agents are fully cooperative -- a common technique to implicitly introduce cooperation among agents in multiagent \gls{rl}.

In our previous work \cite{jadoon2022deep}, we had employed success only reward $R_n(k) = \sum_n G_n(k)$ but such reward, as we have observed, is not enough. Due to limited information availability at each agent and partial observablility, the algorithm doesn't converge well and the performance degrades as the number of agents grows.
\subsection{Multiagent DQN with Parameter Sharing}
In Q-learning, Q-values are used to express the expected reward for each state-action pair as
\begin{equation}   \label{eq:Q}
    Q_n(a,s)
    = \mathsf{E}_\pi\bigl[ \mathcal{R}_n(k) \bigm| A_n(k) = a, S_n(k) = s \bigr]
\end{equation}
where $\mathsf{E}_\pi[\cdot]$ denotes the expectation under the common policy $\pi$, with respect to the current state of the agent.

For the \gls{dqn} \cite{mnih2015humanlevel}, a neural network is used to approximate the Q-values $Q(a,s,\theta) \approx Q^*(s,a)$, where $Q(a,s,\theta)$ is the Q-value estimated by the neural network for action $a$ when the state is $s$, and $\theta$ denotes the weights of the neural network.

In this work, we employ \emph{parameter sharing} method, which basically extends the single agent network to multiple agents \cite{gupta_2017}. The core idea is to use the same function approximator (e.g., neural network) to calculate the Q-values $Q_n(a_n,s_n,\theta)$ for all the agents. Parameter sharing allows us to learn a common policy for all the agents in a centralized way, whilst the deployment of the policy for each agent is in a decentralized manner and therefore, we may drop subscript $n$ from $Q_n(s,a)$.

\begin{figure}[t]
\centering
    \includegraphics[width=250px, height=140px]{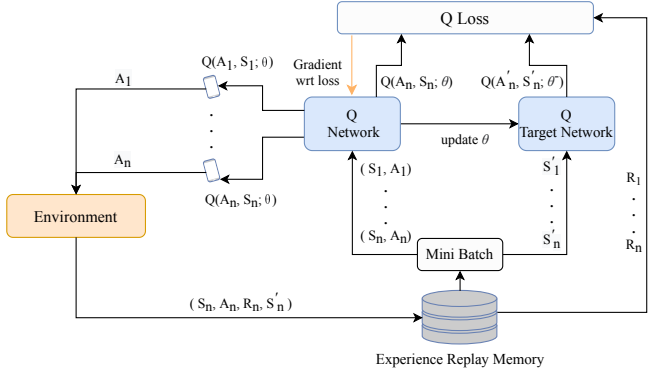}
    \caption{Schematic of the training of the \gls{dqn} with parameter sharing}
    \label{fig:DQN_schematic}
\end{figure}

The parameter sharing method proposed in \cite{gupta_2017} incorporates the IDs of each agent in the state to distinguish between the agents and for each agent to have a unique state every time. In this work, we are employing unsourced \gls{ra} where new agents can join/leave the network any time. Therefore, we are not using any agent/device IDs in the state to distinguish them. 
We use the \textit{experience replay} to the train the \gls{dqn}, which is performed by memorizing the experiences of each agent as $\big(s(k),a(k),r(k),s(k+1)\big)$ in a replay buffer memory$\mathcal{D}$ for each iteration. The learning updates are applied on the experience samples $(s,a,r,s') \sim U(\mathcal{D})$, that are drawn at random with uniform distribution as mini-batches of size $M$ from $\mathcal{D}$. Moreover, we use two neural networks  \cite{doubleDQN}: The Q-network with parameters $\theta$ that is used to evaluate and update the actual policy, and the target network with parameters $\theta^-$. The process is shown in Fig. \ref{fig:DQN_schematic}. After each iteration $i$, the parameters $\theta$ are updated minimizing the following loss function,
\begin{align*}
    L_i(\theta_i) = \mathsf{E}_{(s,a,r,s') \sim U(\mathcal{D})} \Big[ \big(y_i
    & - Q(a, s; \theta_i)\big)^2 \Big], 
\end{align*}

where $y_i =r + \gamma \max_{a'} Q(a', s'; \theta_i^-)$ is the target value for the iteration $i$. 

We obtain the following by differentiating the loss function $L_i(\theta_i)$ with respect to the weights,
\begin{align}
    \nabla_{\theta_i} L_i(\theta_i) = \mathsf{E}_{(s,a,r,s')} \Big[ \big( y_i
    - Q(a, s; \theta_i)\big)  \nabla_{\theta_i}Q(a, s; \theta_i)  \Big] \nonumber
\end{align}
For DQN-RA, at each time slot $k$, each \gls{mtd} $n$ obtains the observation (feedback) $F(k)$ after taking the action $A_n(k) \in \mathcal{A}$; it then updates its history $S_n(k)$ and feeds $S_n(k)$ to the \gls{dqn} as input. The output of the \gls{dqn} the Q-values corresponding to each action. The device $n$ then follows the policy $\pi$ by drawing an action $A_n(k)$ from the following distribution calculated using the softmax policy \cite{Sutton_2E}
\begin{align}
    \pi(a_n|s_n)
    &= \frac{e^{\beta Q(a_n,s_n)}} {\sum\nolimits_{\tilde{a_n} \in \mathcal{A}} e^{\beta Q(\tilde{a}_n,s_n)}} + \frac{\epsilon}{|\mathcal{A}|},
    \qquad \forall a_n \in \mathcal{A},
    \label{eq:strategy}
\end{align}
where $\beta > 0$ is the temperature parameter and $0<\epsilon < 1$, which are used to adjust the balance between \textit{exploration} and \textit{exploitation}.

\begin{algorithm}
\SetAlgoLined
\textbf{Define} $\alpha \in (0, 1]$, $\gamma \in [0, 1]$, $\epsilon > 0$ and number of  \glspl{mtd} $N$ \\
\textbf{Initialize} $S_n(k) = \boldsymbol{0}$, $B_n(k) = 0$ $\forall n \in \mathcal{N}$, weight update frequency $L$, $\lambda_n$, history size $h$ and  total time slots $K = 4\lambda_n N$ \\
\For{each episode} 
    {
    \textbf{Activate} $N_a$ new \glspl{mtd}, i.e., $N_
    a \sim \textrm{Poisson}(\lambda_n N)$ $\forall n \in \mathcal{N}$ \\
    \textbf{Set} $B_n(k)=1 \forall n \in \mathcal{N}_a$ \\  
    \For{$k = 1, \dotsc , K$} 
        {
            \For{each \gls{mtd} $n = 1, \dotsc , N$}
                {
                \If {$B_n(k) \neq 0$}
                    {
                    Observe $S_n(k)$ as in \eqref{eq:history_user_n} and feed it to the Q-network \\
                    Generate the estimate of $Q(a_n)$ $\forall a_n \in \mathcal{A}$ \\
                    Take action $A_n(k)$ according to \eqref{eq:strategy} \\
                    Obtain feedback $F(k)$ as observation and calculate reward $R_n(k)$ \\
                    Update the buffer $B_n(k)$ and obtain the next state $S'_n(k)$ \\
                    Feed $S'_n(k)$ to both Q-network, and target Q-network \\
                    Generate estimates from both Q-networks, $Q(a_n)$ and $Q_\mathrm{target}(a_n)$
                    }
                 }
                   
            Train Q-network with minibatch of size $M$ as input $S(k) = S_1(k), S_2(k), \dotsc, S_M(k)$ and output $M$ Q-values \\ 
            \If{$t \% L = 0$}
                    {
                    $Q_\mathrm{target} \leftarrow Q$
                    }
        }
        Reset $B_n(k) = 0$ $\forall n \in \mathcal{N}$
    }    
    \caption{Training of the proposed DQN-RA}
    \label{alg:DQN_training}
\end{algorithm}

\begin{algorithm}
\SetAlgoLined
\textbf{Initialize} $S_n(k) = \boldsymbol{0}$, $B_n(k) = 0$ $\forall n \in \mathcal{N}$, $\lambda_n$,  and  total time slots $K = 4\lambda_n N$ \\
\textbf{Define} Number of  \glspl{mtd} $N$ and history size $h$ \\
\For{each episode} 
    {
    \textbf{Activate} $N_a$ new \glspl{mtd}, i.e., $N_
    a \sim \textrm{Poisson}(\lambda_n N)$ $\forall n \in \mathcal{N}$ \\
    \textbf{Set} $B_n(k)=1 \forall n \in \mathcal{N}_a$ \\
    \For{$k = 1, \dotsc , K$} 
        {
            \For{each \gls{mtd} $n = 1, \dotsc , N$}
                { 
                \If {$B_n(k) \neq 0$}
                    {
                    Observe $S_n(k)$ as in \eqref{eq:history_user_n} and feed it to Q-network \\
                    Generate the estimate of $Q(a_n)$ $ \forall a_n \in \mathcal{A}$ \\
                    Take action $A_n(k)$ according to \eqref{eq:strategy} \\
                    Obtain feedback $F(k)$\\
                    Update the buffer $B_n(k)$ and the next state $S'_n(k)$ \\
                    }
                } 
        }
        Reset $B_n(k) = 0$ $\forall n \in \mathcal{N}$
    }    
    \caption{Testing phase of the DQN-RA}
    \label{alg:DQN_testing}
\end{algorithm}

\begin{table}[t!]
\centering
\caption{Simulation Paramaters}
\begin{tabular}{|ll|}
\hline
\textbf{Parameter}                             & \textbf{Value}    
 \\ \hline
$\rho$       & 0.2      \\ \hline
$\lambda_n$       & 0.05      \\ \hline
$(p_{\mathrm{min}}, p_{\mathrm{max}})$ for EB schemes & $(0.001, 0.9)$      \\ \hline
Total time slots $K$          & $4\lambda_nN$                            \\ \hline
History size $h$                      & 5                          \\ \hline
Learning rate  & 1e-4                         \\ \hline
$(\epsilon, \epsilon_{\mathrm{min}})$ & $(0.5, 0.1)$  \\ \hline
Temperature $\beta$ & $1 - 15$  \\ \hline
$\#$ of hidden layers & 2, (150 and 100 units) \\ \hline
Batch size         & 8 \\ \hline
\# of episodes (training \& testing)          & 50                            \\ \hline
\end{tabular}
\label{tab:simulation}
\end{table}

\section{Simulation Results and Discussion}
The \gls{dqn} contain two fully connected hidden layers with $150$ and $100$ units each. We employ episodic training to produce the results. At the start of each episode, out of $N$ \glspl{mtd}, on average $N_a$ devices become active following the random process, i.e., $N_a \sim \mathrm{Poisson}(\lambda_nN)$. Each \gls{mtd} $n \in \mathcal{N}_a$ has one packet in its buffer, i.e, $B_n(k) = 1 \forall n \in \mathcal{N}_a$. Each episode comprises of $K$ time slots and $K$ depends on the number of  \glspl{mtd} and $\lambda_n$. For our results we use $K = 4\lambda_nN$, which allows enough time slots for both \gls{dqn} and \gls{eb} policies to resolve the collisions and successfully transmit their packets. Since the average arrival rate for each device $\lambda_n$ remains the same; however due to the fact that the total arrival rate of the system is $\lambda_n N$, as the $N$ grows, the total number of devices becoming active also grow.

The training and testing process of DQN-RA is depicted in Algorithm. \ref{alg:DQN_training} and Algorithm. \ref{alg:DQN_testing} respectively. The parameters used in the episodic training of the \gls{dqn} and also to generate the simulation results are given in Table. \ref{tab:simulation}. Please note that only devices that are active, i.e., $B_n(k) \neq 0$ are passed through the Q-networks. For inactive devices, we use zero-masking where the value of the state is set to $0$ values. The states of inactive agents is still used to update the experience replay buffer.  
\begin{figure}[t!]
\centering
    \includegraphics[width=240px, height=160px]{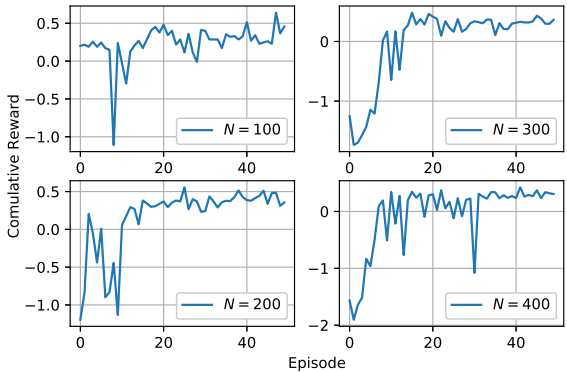}
    \caption{Average cumulative reward for different values of $N$.}
    \label{fig:rewards_N}
\end{figure}

In Fig. \ref{fig:rewards_N} we show the reward trends during the training of the \gls{dqn} for different values of $N$. For space constraints, we are not showing for all the values of $N$ that we have used but they all converge in a similar manner. Small fluctuations are due to the randomness as the number of devices becoming active is not constant or the same for each episode. Next we show the performance of our proposed schemes in terms of throughput, delay and collision rate and we compare the performance with binary \gls{eb} schemes, both BnSEB and BSEB. 

\begin{figure}[t]
\centering
    \includegraphics[width=240px, height=160px]{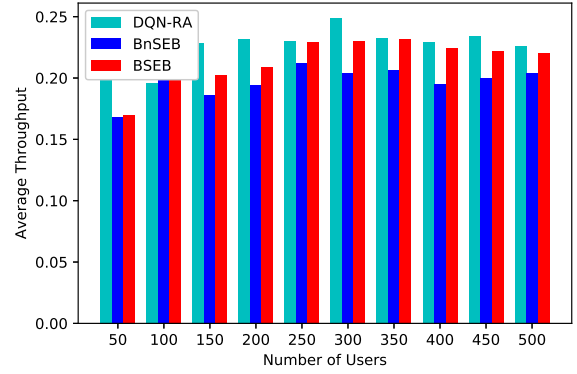}
    \caption{Average throughput comparison.}
    \label{fig:throughput_ep}
\end{figure}

\begin{figure}[t]
\centering
    \includegraphics[width=240px, height=160px]{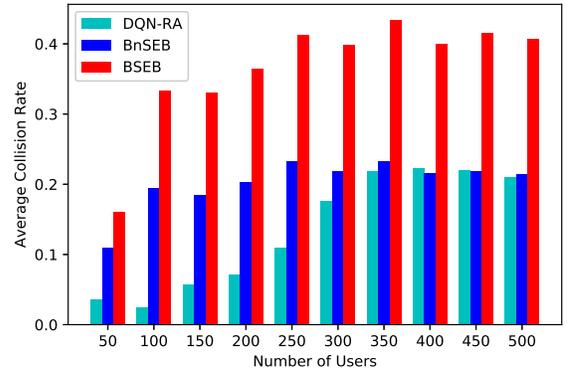}
    \caption{Average collision rate comparison.}
    \label{fig:pcr_ep}
\end{figure}
The average throughput of the system and the average packet collision rate are shown in Fig. \ref{fig:throughput_ep} and Fig. \ref{fig:pcr_ep} respectively. Comparing both Fig. \ref{fig:throughput_ep} and Fig. \ref{fig:pcr_ep} it is clear that the learned DQN-RA policy achieves better throughput compared to both BnSEB and BSEB with $\sigma = 2$ whilst having lower collision rate. Moreover, in Fig. \ref{fig:throughput_ep}, BSEB achieves better throughput as compared to the BnSEB because the devices transmit more aggressively when they increase and decrease their transmit probabilities together and be able to achieve better throughput. However, due to this behavior, BSEB has highest collision rate as compared to the BnSEB as depicted in Fig. \ref{fig:pcr_ep}. Clearly, DQN-RA finds a balance between both approaches and outperforms both BnSEB and BSEB both in terms of average throughput and collision rate. The DQN-RA has similar performance to BnSEB for higher number of \glspl{mtd} in terms of packet collision rate but it exhibits lower average packet delay even for higher number of devices.

\begin{figure}[t]
\centering
    \includegraphics[width=240px, height=160px]{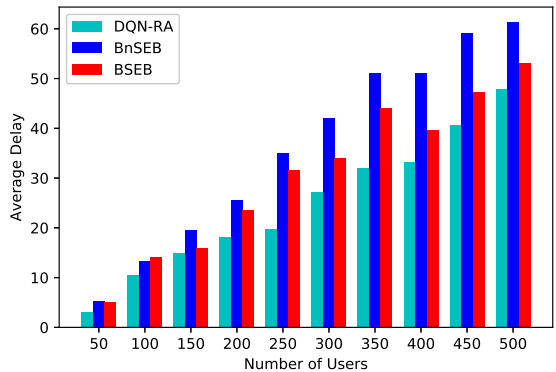}
    \caption{Average delay comparison.}
    \label{fig:delay_ep}
\end{figure}

Similarly, Fig. \ref{fig:delay_ep} shows the performance of average delay and we can observe that the proposed approach incurs lowest average packet delay as compared to both BnSEB and BSEB as the number of \ grow. BnSEB has the highest average delay because packets stay in the buffer due to the fact that devices significantly reduce their transmit probabilities and therefore it is also reflected in the behaviour of this scheme in Fig. \ref{fig:throughput_ep} which shows that  \glspl{mtd} are unable to transmit frequently even when channel is free and in  Fig. \ref{fig:pcr_ep} it has therefore the low collision rate. It becomes more apparent when the value $N$ becomes higher. The proposed scheme outperforms both \gls{eb} techniques even in terms of average packet delay.

We set the history size $h = 5$ for all the experiments for DQN-RA and the performance for each value of $N$ is calculated as the average over $50$ episodes for all the schemes. We have also tried the experiments with $h=1,3, 10$ but history size $h=5$ performs better and increasing $h$ does not further improve the performance.  Due to space constraints we are not showing the results here. Furthermore, the temperature parameter $\beta$ and $\epsilon$ are used for exploration and we increase $\beta$ and decrease $\epsilon$ during the training for each $N$. The $\epsilon_{\mathrm{min}} = 0.1$, which is kept at this value to prevent the transmit probabilities (policy) to go to $0$ when devices start colliding at the start of each episode. Therefore, the  value of $\epsilon_{\mathrm{min}}$ puts a lower bound on the transmission probability of each \gls{mtd} for stability. 

\section{Conclusion}
In this work, we propose a collision resolution policy for \gls{ra} in \gls{mtc} where the devices can become active and inactive randomly. We provide the performance comparison of our proposed DQN-RA policy with \gls{eb} schemes and show that our proposed policy performs better in terms of average throughput, collision rate and delay. We use parameter sharing method with \gls{dqn} to learn a single policy that is learned in a centralized manner and it can be executed distributively by every \gls{mtd}. We show that our scheme scales well for higher number of  \glspl{mtd}. In our next work, we will use different traffic arrival methods suitable for \gls{mtc}, e.g., the ones mentioned in \cite{navarro20}, and we will explore other multi-agent \gls{rl} algorithms such as policy gradient methods to learn the transmit probabilities of the devices. Furthermore, for \gls{mtc} traffic, exploiting the advantages of both scheduled access and \gls{ra} might be a better way to manage massive access, which we will also explore in our future work.

\section*{Acknowledgment}
The work of A. Pastore and M. Navarro was supported by Grant RTI2018-099722-B-I00 funded by MCIN/AEI/10.13039/501100011033 and by ``ERDF A way of making Europe''. The work of M. A. Jadoon was supported by the European Union H2020 Research and Innovation Programme through Marie Skłodowska Curie action (MSCA-ITN-ETN 813999 WINDMILL).

\bibliographystyle{ieeetr}
\bibliography{bibliography}

\begin{thebibliography}{10}

\bibitem{bockelmann16}
C.~Bockelmann, N.~Pratas, H.~Nikopour, K.~Au, T.~Svensson, C.~Stefanovic,
  P.~Popovski, and A.~Dekorsy, ``Massive machine-type communications in 5{G}:
  physical and {MAC}-layer solutions,'' {\em IEEE Communications Magazine},
  vol.~54, no.~9, pp.~59--65, 2016.

\bibitem{Barletta18}
L.~Barletta, F.~Borgonovo, and I.~Filippini, ``The throughput and access delay
  of slotted-aloha with exponential backoff,'' {\em IEEE/ACM Transactions on
  Networking}, vol.~26, no.~1, pp.~451--464, 2018.

\bibitem{CHU201523}
Y.~Chu, S.~Kosunalp, P.~D. Mitchell, D.~Grace, and T.~Clarke, ``Application of
  reinforcement learning to medium access control for wireless sensor
  networks,'' {\em Engineering Applications of Artificial Intelligence},
  vol.~46, pp.~23--32, 2015.

\bibitem{Naparstek}
O.~{Naparstek} and K.~{Cohen}, ``Deep multi-user reinforcement learning for
  distributed dynamic spectrum access,'' {\em IEEE Transactions on Wireless
  Communications}, vol.~18, no.~1, pp.~310--323, 2019.

\bibitem{zhong2019deep}
C.~Zhong, Z.~Lu, M.~C. Gursoy, and S.~Velipasalar, ``Actor-critic deep
  reinforcement learning for dynamic multichannel access,'' in {\em 2018 IEEE
  Global Conference on Signal and Information Processing (GlobalSIP)},
  pp.~599--603, 2018.

\bibitem{wang2018}
S.~Wang, H.~Liu, P.~H. Gomes, and B.~Krishnamachari, ``Deep reinforcement
  learning for dynamic multichannel access in wireless networks,'' {\em IEEE
  Transactions on Cognitive Communications and Networking}, vol.~4, no.~2,
  pp.~257--265, 2018.

\bibitem{Tomovic20}
S.~Tomovic and I.~Radusinovic, ``A novel deep {Q}-learning method for dynamic
  spectrum access,'' in {\em 2020 28th Telecommunications Forum (TELFOR)},
  pp.~1--4, 2020.

\bibitem{alfaro_alohaq}
L.~{de Alfaro}, M.~{Zhang}, and J.~J. {Garcia-Luna-Aceves}, ``Approaching fair
  collision-free channel access with slotted aloha using collaborative
  policy-based reinforcement learning,'' in {\em 2020 IFIP Networking
  Conference (Networking)}, pp.~262--270, 2020.

\bibitem{Yang21}
H.~Yang, Z.~Xiong, J.~Zhao, D.~Niyato, C.~Yuen, and R.~Deng, ``Deep
  reinforcement learning based massive access management for ultra-reliable
  low-latency communications,'' {\em IEEE Transactions on Wireless
  Communications}, vol.~20, no.~5, pp.~2977--2990, 2021.

\bibitem{jadoon2022deep}
M.~A. Jadoon, A.~Pastore, M.~Navarro, and F.~Perez-Cruz, ``Deep reinforcement
  learning for random access in machine-type communication,'' in {\em 2022 IEEE
  Wireless Communications and Networking Conference (WCNC)}, 2022.

\bibitem{mnih2015humanlevel}
Mnih {\em et~al.}, ``Human-level control through deep reinforcement learning,''
  {\em Nature}, vol.~518, pp.~529--533, Feb. 2015.

\bibitem{gupta_2017}
J.~K. Gupta, M.~Egorov, and M.~Kochenderfer, ``Cooperative multi-agent control
  using deep reinforcement learning,'' in {\em Autonomous Agents and Multiagent
  Systems} (G.~Sukthankar and J.~A. Rodriguez-Aguilar, eds.), (Cham),
  pp.~66--83, Springer International Publishing, 2017.

\bibitem{doubleDQN}
H.~van Hasselt, A.~Guez, and D.~Silver, ``Deep reinforcement learning with
  double {Q}-learning,'' 2015.

\bibitem{Sutton_2E}
R.~S. Sutton and A.~G. Barto, {\em Reinforcement Learning: An Introduction.
  (Second Edition)}.
\newblock Cambridge, MA, USA: A Bradford Book, 2018.

\bibitem{navarro20}
J.~Navarro-Ortiz, P.~Romero-Diaz, S.~Sendra, P.~Ameigeiras, J.~J. Ramos-Munoz,
  and J.~M. Lopez-Soler, ``A survey on 5g usage scenarios and traffic models,''
  {\em IEEE Communications Surveys Tutorials}, vol.~22, no.~2, pp.~905--929,
  2020.

\end{thebibliography}

\end{document}